\documentclass[aps,preprint,showpacs,preprintnumbers,eqsecnum,amsmath,amssymb,11pt]{revtex4}

\usepackage{graphics,epsfig}
\usepackage{graphicx}
\usepackage{dcolumn}
\usepackage{bm}

\begin{document}

\title{Degradation of non-maximal entanglement of scalar and
Dirac fields in non-inertial frames}
\author{Qiyuan Pan \ \ \ \ Jiliang  Jing }
\thanks{Corresponding author, Electronic address:
jljing@hunnu.edu.cn} \affiliation{Institute of Physics and
Department of Physics, Hunan Normal University, Changsha, Hunan
410081, P. R. China\\ and
\\ Key Laboratory of Low Dimensional Quantum Structures and
Quantum Control of Ministry of Education, Hunan Normal University ,
Changsha, Hunan 410081, P.R. China}


\begin{abstract}

The entanglement between two modes of the free scalar and Dirac
fields as seen by two relatively accelerated observers has been
investigated. It is found that the same initial entanglement for an
initial state parameter $\alpha$ and its ``normalized partner"
$\sqrt{1-\alpha^{2}}$ will be degraded by the Unruh effect along two
different trajectories except for the maximally entangled state,
which just shows the inequivalence of the quantization for a free
field in the Minkowski and Rindler coordinates. In the infinite
acceleration limit the state doesn't have the distillable
entanglement for any $\alpha$ for the scalar field but always
remains entangled to a degree which is dependent of $\alpha$ for the
Dirac field. It is also interesting to note that in this limit the
mutual information equals to just half of the initially mutual
information, which is independent of $\alpha$ and the type of field.

\end{abstract}

 \pacs{03.65.Ud, 03.67.Mn, 04.70.Dy, 97.60.Lf}

\maketitle

The quantum information theory has made rapid progress in recent
years and more and more efforts have been expended on the study of
quantum information in the relativistic framework \cite{PT-B-P-E}.
Especially, the entanglement in a relativistic setting has received
considerable attention because it is considered to be a major
resource for quantum information tasks such as quantum
teleportation, quantum computation and so on \cite{DAA}. Despite the
potential interest to quantum information, the study of entanglement
can also help us get a deeper understanding of the black hole
thermodynamics \cite{Bombelli-Callen} and the black hole information
paradox \cite{Hawking-Terashima}. Thus, many authors have
investigated the entanglement in the relativistic frames inertial or
not for various fields \cite{PST,Mann,Alsing}.

More recently, Fuentes-Schuller \emph{et al.} \cite{Mann} and Alsing
\emph{et al.} \cite{Alsing} explicitly demonstrated that the
entanglement is a quantity depending on a relative acceleration of
one of the observers who, before being accelerated, shared a
maximally entangled bosonic or fermionic pair. Their results also
showed that the different type of field will have a qualitatively
different effect on the degradation of entanglement produced by the
Unruh effect \cite{Davies-Unruh}. Choosing a generic state as the
initial entangled state in this Brief Report
\begin{eqnarray}\label{initial}
|\Psi_{sk}\rangle=\sqrt{1-\alpha^{2}}|0_{s}\rangle^{M}|1_{k}\rangle^{M}
+\alpha|1_{s}\rangle^{M}|0_{k}\rangle^{M},
\end{eqnarray}
where $\alpha$ is some real number which satisfies
$|\alpha|\in(0,1)$, $\alpha$ and $\sqrt{1-\alpha^{2}}$ are the
so-called ``normalized partners", we will try to see what effects
this uncertainly initial entangled state will on the degradation of
entanglement for two relatively accelerated observers due to the
presence of an initial state parameter $\alpha$. Notice that the
Schwarzschild space-time very close to the event horizon resembles
the Rindler space in the infinite acceleration limit
\cite{Wald,Mann}. Hence, as in \cite{Mann,Alsing} our results in
this limit can be applied to discuss the entanglement between two
free bosonic or fermionic modes seen by observers when one observer
falls into a black hole and the other barely escapes through eternal
uniform acceleration.

Rindler coordinates are appropriate for describing the viewpoint of
an observer moving with uniform acceleration. The world lines of
uniformly accelerated observers in the Minkowski coordinates
correspond to hyperbolae in the left (region I) and right (region
II) of the origin which are bounded by light-like asymptotes
constituting the Rindler horizon \cite{Mann,Alsing}, so two Rindler
regions are causally disconnected from each other \cite{Birrell}. An
observer undergoing uniform acceleration remains constrained to
either Rindler region I or II and has no access to the other sector.
The system in Eq. (\ref{initial}) is bipartite from an inertial
perspective, but in a non-inertial frame an extra set of modes in
region II becomes relevant. Thus, we will study the mixed-state
entanglement of the state as seen by an inertial observer Alice
detecting the mode $s$ and a uniformly accelerated observer Bob with
proper acceleration $a$ in region I detecting the second mode $k$.

\begin{figure}[ht]
\includegraphics[scale=0.7]{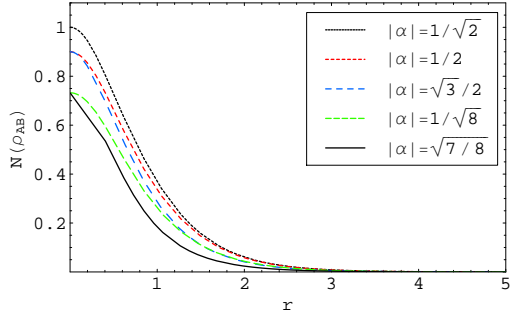}\vspace{0.0cm}
\caption{\label{NegBos} Logarithmic negativity of the bosonic modes
versus $r$ for different $\alpha$.}
\end{figure}

\textit{Bosonic entanglement} For a free scalar field, the Minkowski
vacuum state can be expressed as a two-mode squeezed state in the
Rindler frame \cite{Davies-Unruh,Birrell}
\begin{eqnarray} \label{Scalar-vacuum}
|0_{k}\rangle^{M}=\frac{1}{\cosh
r}\sum_{n=0}^{\infty}\tanh^{n}r|n_{k}\rangle_{I}|n_{k}\rangle_{II},
\end{eqnarray}
where $\cosh r=(1-e^{-2\pi|k|c/a})^{-1/2}$, $k$ is the wave vector
and $r$ is the acceleration parameter, $|n\rangle_{I}$ and
$|n\rangle_{II}$ indicate the Rindler-region-I-particle mode and
-II-antiparticle mode respectively. Using Eq. (\ref{Scalar-vacuum})
and the first excited state \cite{Mann,Birrell}
\begin{eqnarray*}\label{Scalar-excited}
|1_{k}\rangle^{M}=\frac{1}{\cosh^{2}
r}\sum_{n=0}^{\infty}\tanh^{n}r\sqrt{n+1}
|(n+1)_{k}\rangle_{I}|n_{k}\rangle_{II},
\end{eqnarray*}
we can rewrite Eq. (\ref{initial}) in terms of Minkowski modes for
Alice and Rindler modes for Bob. Since Bob is causally disconnected
from region II, we will trace over the states in this region and
obtain
\begin{eqnarray}\label{Bos-density}
&&\rho_{AB}=\frac{1}{\cosh^{2}r}\sum_{n=0}^{\infty}\tanh^{2n}r\rho_{n},
\nonumber\\&& \rho_{n}=\alpha^{2}|1n\rangle\langle1n|
+\frac{\alpha\sqrt{(1-\alpha^{2})(n+1)}}{\cosh
r}|1n\rangle\langle0(n+1)| \nonumber\\&&\qquad
+\frac{\alpha\sqrt{(1-\alpha^{2})(n+1)}}{\cosh
r}|0(n+1)\rangle\langle1n| \nonumber\\&&\qquad
+\frac{(1-\alpha^{2})(n+1)}{\cosh^{2}
r}|0(n+1)\rangle\langle0(n+1)|,
\end{eqnarray}
where $|nm\rangle=|n_{s}\rangle^{M}|m_{k}\rangle_{I}$. The partial
transpose criterion provides a sufficient condition for the
existence of entanglement in this case \cite{peres}: if at least one
eigenvalue of the partial transpose is negative, the density matrix
is entangled; but a state with positive partial transpose can still
be entangled. It is well-known bound or nondistillable entanglement
\cite{Vidal-Plenio}. Interchanging Alice's qubits, we get the
eigenvalues of the partial transpose $\rho_{AB}^{T_{A}}$ in the
($n$,$n+1$) block
\begin{eqnarray*}
\lambda_{\pm }^{n}=\frac{\tanh^{2n}r}{2\cosh ^{2}r }
\left[\xi_n\pm\sqrt{\xi_{n}^{2}+\frac{4\alpha^{2}
(1-\alpha^{2})}{\cosh^{2}r}}\right],
\end{eqnarray*}
where $\xi_{n}=\alpha^{2}\tanh^{2}r+(1-\alpha^{2})n/\sinh^{2}r$.
Obviously the eigenvalue $\lambda_{-}^{n}$ is always negative for
finite acceleration ($r<\infty$). Hence, this mixed state is always
entangled for any finite acceleration of Bob. In the limit
$r\rightarrow\infty$, the negative eigenvalue will go to zero. To
discuss this further, we will use the logarithmic negativity which
serves as an upper bound on the entanglement of distillation
\cite{Vidal-Plenio}. This entanglement monotone is defined as
$N(\rho )=\log _{2}||\rho^{T}||$, where $||\rho^{T}||$ is the trace
norm of the partial transpose $\rho^{T}$. We therefore find
\begin{eqnarray*}
&&N(\rho_{AB})=\log_{2}\left\{\frac{\alpha^{2}}{\cosh^{2}{r}}
+\sum_{n=0}^{\infty}\frac{\tanh^{2n}r}{\cosh^{2}r}\right.
\nonumber\\&&\times\left.\sqrt{\left[\alpha^{2}\tanh^{2}r
+\frac{(1-\alpha^{2})n}{\sinh^{2}r}\right]^{2} +\frac{4\alpha^{2}
(1-\alpha^{2})}{\cosh^{2}r}}\right\}.
\end{eqnarray*}
For vanishing acceleration ($r=0$),
$N(\rho_{AB})=\log_{2}(1+2|\alpha|\sqrt{1-\alpha^{2}})$. In the
range $0<|\alpha|\leq1/\sqrt{2}$ the larger $\alpha$, the stronger
the initial entanglement; but in the range
$1/\sqrt{2}\leq|\alpha|<1$ the larger $\alpha$, the weaker the
initial entanglement. For finite acceleration, the monotonous
decrease of $N(\rho_{AB})$ with increasing $r$ for different
$\alpha$ means that the entanglement of the initial state is lost to
the thermal fields generated by the Unruh effect. From Fig.
\ref{NegBos} it is surprisingly found that the same initial
entanglement for $\alpha$ and its ``normalized partner"
$\sqrt{1-\alpha^{2}}$ will be degraded along two different
trajectories except for the maximally entangled state, i.e.,
$|\alpha|=1/\sqrt{2}$. This phenomenon, due to the coupling of
$\alpha$ and the hyperbolic functions related to $r$, just shows the
inequivalence of the quantization for a scalar field in the
Minkowski and Rindler coordinates. The logarithmic negativity is
exactly zero for any $\alpha$ in the limit $r\rightarrow\infty$,
which indicates that the state doesn't have the distillable
entanglement.

\begin{figure}[ht]
\includegraphics[scale=0.7]{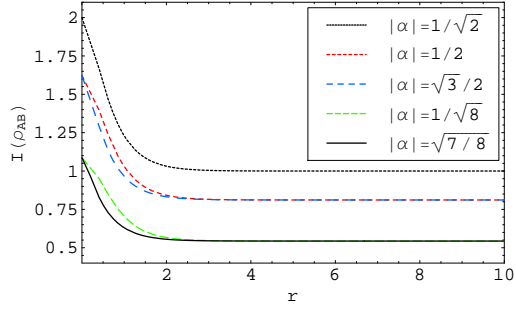} \vspace{0.0cm}
\caption{\label{MuInBos} Mutual information of the bosonic modes
versus $r$ for different $\alpha$.}
\end{figure}

The mutual information, which can be used to estimate the total
amount of correlations between any two subsystem of the overall
system, is defined as \cite{RAM}
\begin{eqnarray}
I(\rho_{AB})=S(\rho_{A})+S(\rho_{B})-S(\rho_{AB}),
\end{eqnarray}
where $S(\rho )=-\text{Tr}(\rho \log_{2}\rho )$ is the entropy of
the density matrix $\rho$. From Eq. (\ref{Bos-density}), we can
obtain the entropy of this joint state
\begin{eqnarray}
&&S(\rho _{AB})=-\sum_{n=0}^{\infty}\frac{\tanh^{2n}r}{\cosh
^{2}r}\left[\alpha^{2}+\frac{(1-\alpha^{2})(n+1)}{\cosh^{2}r}\right]
\nonumber\\&&\qquad\times\log_{2}\frac{{\tanh ^{2n}r}}{\cosh
^{2}r}\left[\alpha^{2}+\frac{(1-\alpha^{2})(n+1)}{\cosh
^{2}r}\right].
\end{eqnarray}
Tracing over Alice's states for $\rho_{AB}$, we get Bob's density
matrix in region I; its entropy is
\begin{eqnarray}
&&S(\rho_{BI})=-\sum_{n=0}^{\infty }\frac{\tanh^{2n}r}{\cosh
^{2}r}\left[\alpha^{2}+\frac{(1-\alpha^{2})n}{\sinh^{2}r}\right]
\nonumber\\&&\qquad\times\log_{2}\frac{\tanh^{2n}r}
{\cosh^{2}r}\left[\alpha^{2}+\frac{(1-\alpha^{2})n}{\sinh^{2}r}\right].
\end{eqnarray}
In the same way, we have Alice's density matrix by tracing over
Bob's states; its entropy is given by
\begin{eqnarray}\label{Alice-entropy}
S(\rho_{A})=-[\alpha^{2}\log_{2}\alpha^{2}
+(1-\alpha^{2})\log_{2}(1-\alpha^{2})].
\end{eqnarray}
We draw the behaviors of the mutual information $I(\rho_{AB})$
versus $r$ for different $\alpha$ in Fig. \ref{MuInBos}. For
vanishing acceleration, the initially mutual information is
$I_{bi}=-2[\alpha^{2}\log_{2}\alpha^{2}
+(1-\alpha^{2})\log_{2}(1-\alpha^{2})]$. In the range
$0<|\alpha|\leq1/\sqrt{2}$ the larger $\alpha$, the stronger
$I_{bi}$; but in the range $1/\sqrt{2}\leq|\alpha|<1$ the larger
$\alpha$, the weaker $I_{bi}$. As the acceleration increases, the
mutual information becomes smaller. It is interesting to note that
except for the maximally entangled state, the same initially mutual
information for $\alpha$ and $\sqrt{1-\alpha^{2}}$ will be degraded
along two different trajectories. However, in the infinite
acceleration limit, the mutual information converges to the same
value again, i.e., $I_{bf}=-[\alpha^{2}\log_{2}\alpha^{2}
+(1-\alpha^{2})\log_{2}(1-\alpha^{2})]$, which equals to just half
of $I_{bi}$. Obviously if $I_{bi}$ is higher, it is degraded to a
higher degree in this limit. Since the distillable entanglement in
the infinite acceleration limit is zero, we are safe to say that the
total correlations consist of classical correlations plus bound
entanglement in this limit.

\textit{Fermionic entanglement} With the single-mode approximation
used by Alsing \emph{et al.}, the fermionic Minkowski vacuum can be
written as \cite{Alsing}
\begin{eqnarray}\label{Dirac-vacuum}
|0\rangle ^{M}=\cos r|0\rangle_{I}|0\rangle _{II}+\sin
r|1\rangle_{I}|1\rangle _{II},
\end{eqnarray}
and the only excited state is given by
\begin{eqnarray}\label{Dirac-excited}
&&|1\rangle ^{M}= |1\rangle_{I}|0\rangle _{II},
\end{eqnarray}
where $\cos r=(1+e^{-2\pi\omega c/a})^{-1/2}$ and the acceleration
parameter $r$ is in the range $0\leq r\leq\pi/4$ for $0\leq
a\leq\infty$ in this case. Using Eq. (\ref{Dirac-vacuum}) and
(\ref{Dirac-excited}) for the Minkowski particle states
$|0_{k}\rangle^{M}$ and $|1_{k}\rangle^{M}$ and tracing over the
modes in the region II, we get
\begin{eqnarray}\label{DM11}
&&\rho_{AB}=(1-\alpha^{2})|01\rangle\langle01|
\nonumber\\&&\qquad~+\alpha\sqrt{(1-\alpha^{2})}\cos
r(|01\rangle\langle10|+|10\rangle\langle01|)
\nonumber\\&&\qquad~+\alpha^{2}(\cos^{2}r|10\rangle\langle10|
+\sin^{2}r|11\rangle\langle11|),
\end{eqnarray}
with $|mn\rangle=|m\rangle_{A}^{M}|n\rangle_{BI}$. The partial
transpose criterion provides a necessary and sufficient condition
for entanglement in a mixed state of two qubits \cite{peres}: if at
least one eigenvalue of the partial transpose is negative, the
density matrix is entangled. Interchanging Alice's qubits, we obtain
an eigenvalue of the partial transpose $\rho_{AB}^{T_{A}}$
\begin{eqnarray*}
\lambda_{-}=\frac{1}{2}\left[\alpha^{2}\sin^{2}{r}
-\sqrt{\alpha^{4}\sin^{4}{r}+4\alpha^{2}(1-\alpha^{2})\cos^{2}{r}}\right],
\end{eqnarray*}
which is always negative for $0\leq r\leq \pi/4$. Thus, the state is
always entangled for any uniform acceleration of Bob. The
logarithmic negativity is expressed as
\begin{eqnarray*}
&&N(\rho_{AB})=\log_{2}[1-\alpha^{2}\sin^{2}{r}
\nonumber\\&&\quad\quad
+\sqrt{\alpha^{4}\sin^{4}{r}+4\alpha^{2}(1-\alpha^{2})\cos^{2}{r}}~].
\end{eqnarray*}
\begin{figure}[ht]
\includegraphics[scale=0.7]{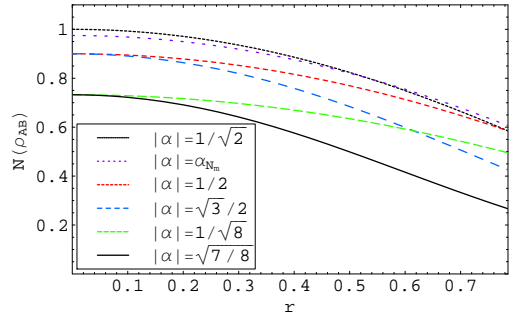} \vspace{0.0cm}
\caption{\label{NegFer} Logarithmic negativity of the fermionic
modes versus $r$ for different $\alpha$ (notice that
$\alpha_{N_{m}}=\sqrt{(4-\sqrt{2})/7}$~).}
\end{figure}
For vanishing acceleration ($r=0$),
$N(\rho_{AB})=\log_{2}(1+2|\alpha|\sqrt{1-\alpha^{2}})$. For finite
acceleration, the entanglement is degraded by the Unruh effect just
as shown in Fig. \ref{NegFer}. We find that in the range
$0<|\alpha|\leq1/\sqrt{2}$ for the larger $\alpha$, the initial
entanglement is higher, but it isn't always degraded to a higher
degree. It should be noted that for $1/2<|\alpha|<1/\sqrt{2}$ the
final entanglement of the initial state is higher than that of the
maximally entangled state, i.e., $\log_{2}(3/2)\simeq0.585$, and for
$|\alpha|=\sqrt{(4-\sqrt{2})/7}$ the maximally final entanglement is
$\log_{2}[(5+4\sqrt{2})/7]\simeq0.606$. In the range
$1/\sqrt{2}\leq|\alpha|<1$ the larger $\alpha$, the weaker the
initial entanglement and the lower the final entanglement. Unlike
the behaviors of the bosonic case, except for the maximally
entangled state, the same initial entanglement of the fermionic
modes for $\alpha$ and $\sqrt{1-\alpha^{2}}$ will be degraded along
two different trajectories and asymptotically reach two differently
nonvanishing minimum values in the infinite acceleration limit
($r=\pi/4$) due to the coupling of $\alpha$ and the trigonometric
functions related to $r$. In the infinite acceleration limit
$N(\rho_{AB})=\log_{2}(1-\alpha^{2}/2+|\alpha|\sqrt{2-7\alpha^{2}/4})\neq0$,
which means that the state is always entangled. This is in strong
contrast to the bosonic case and shows that the fermionic system can
be used as a resource for performing certain quantum information
processing tasks.

\begin{figure}[ht]
\includegraphics[scale=0.7]{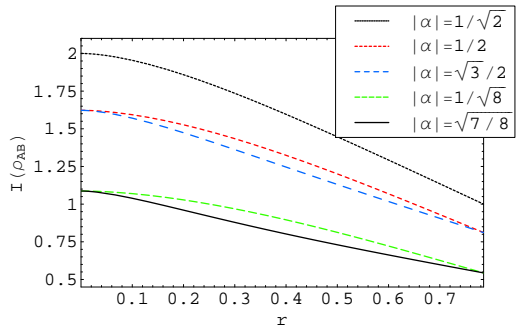}\vspace{0.0cm}
\caption{\label{MuInFer} Mutual information of the fermionic modes
versus $r$ for different $\alpha$.}
\end{figure}

Similar to the bosonic case, we go through the same process again
and get the mutual information for these fermionic modes
\begin{eqnarray}
&&I(\rho_{AB})=(1-\alpha^{2}\sin^{2}{r})\log_{2}(1-\alpha^{2}\sin^{2}{r})
\nonumber\\&&\quad\quad
+\alpha^{2}\sin^{2}{r}\log_{2}\alpha^{2}\sin^{2}{r}
\nonumber\\&&\quad\quad
-(1-\alpha^{2}\cos^{2}{r})\log_{2}(1-\alpha^{2}\cos^{2}{r})
\nonumber\\&&\quad\quad
-\alpha^{2}\cos^{2}{r}\log_{2}\alpha^{2}\cos^{2}{r}
\nonumber\\&&\quad\quad
-\alpha^{2}\log_{2}\alpha^{2}-(1-\alpha^{2})\log_{2}(1-\alpha^{2}),
\end{eqnarray}
whose trajectories versus $r$ for different $\alpha$ are shown by
Fig. \ref{MuInFer}.
For vanishing acceleration, the initially mutual information is
$I_{fi}=-2[\alpha^{2}\log_{2}\alpha^{2}
+(1-\alpha^{2})\log_{2}(1-\alpha^{2})]$, whose behaviors are the
same to $I_{bi}$ of the bosonic modes. The mutual information
becomes smaller as the acceleration increases, and again we
surprisingly find that the same initially mutual information for
$\alpha$ and $\sqrt{1-\alpha^{2}}$ will be degraded along two
different trajectories except for the maximally entangled state. In
the infinite acceleration limit the mutual information converges to
$I_{ff}=-[\alpha^{2}\log_{2}\alpha^{2}
+(1-\alpha^{2})\log_{2}(1-\alpha^{2})]$, which is just half of
$I_{fi}$. This behavior is reminiscent of that seen for the bosonic
case, so we conclude that
\begin{eqnarray}
&&I_{f}=\frac{1}{2}I_{i},
\end{eqnarray}
which is independent of $\alpha$ and the type of field.

It should be noted that if we set the initial entangled state as
\begin{eqnarray}\label{initial-BF}
|\Psi_{sk}\rangle=\alpha|0_{s}\rangle^{M}|0_{k}\rangle^{M}
+\sqrt{1-\alpha^{2}}|1_{s}\rangle^{M}|1_{k}\rangle^{M},
\end{eqnarray}
we will have the same behavior of the entanglement degradation for
the same $\alpha$ just as shown in Figs. \ref{NegBos}-\ref{MuInFer}.

Summarizing, the entanglement of the scalar and Dirac fields in
non-inertial frames is degraded by the Unruh effect as the Bob's
rate of acceleration increases, but their behaviors of the
degradation of entanglement are different for the same initial state
parameter $\alpha$. It is surprisingly found that the same initial
entanglement for $\alpha$ and $\sqrt{1-\alpha^{2}}$ will be degraded
along two different trajectories except for the maximally entangled
state, which just shows the inequivalence of the quantization for a
free field in the Minkowski and Rindler coordinates. In the infinite
acceleration limit, which can be applied to the case Alice falling
into a black hole while Bob barely escapes, the state doesn't have
the distillable entanglement for any $\alpha$ for the scalar field
but always remains entangled to a degree which is dependent of
$\alpha$ for the Dirac field. It should be noted that for
$|\alpha|=\sqrt{(4-\sqrt{2})/7}$, we will have the maximally final
entanglement for the fermionic state in this limit. Further analysis
shows that the mutual information is degraded to a nonvanishing
minimum value which is dependent of $\alpha$ for these two fields
with increasing acceleration parameter $r$. However, it is
interesting to note that the mutual information in the infinite
acceleration limit equals to just half of the initially mutual
information, which is independent of $\alpha$ and the type of field.

This work was supported by the National Natural Science Foundation
of China under Grant No. 10675045; the FANEDD under Grant No.
200317; the Hunan Provincial Natural Science Foundation of China
under Grant No. 07A0128; and the construct program of the key
discipline in hunan province.

\end{document}